**How Well Did U.S. Rail and Intermodal Freight Respond to the COVID-19 Pandemic vs. the Great Recession?**

**Max T.M. Ng**
Graduate Research Assistant
Transportation Center
Northwestern University
600 Foster Street
Evanston, IL 60208, USA
Email: maxng@u.northwestern.edu

**Joseph L. Schofer**
Professor Emeritus of Civil and Environmental Engineering
Northwestern University
2145 Sheridan Road, Evanston, IL 60208, USA
Email: j-schofer@northwestern.edu

**Hani S. Mahmassani***
William A. Patterson Distinguished Chair in Transportation
Director, Transportation Center
Northwestern University
600 Foster Street
Evanston, IL 60208, USA
Email: masmah@northwestern.edu
Tel: 847.491.2276

Submission Date: December 26, 2024

* Corresponding Author

**ABSTRACT**

This paper analyzes and compares patterns of U.S. domestic rail freight volumes during, and after the disruptions caused by the 2007-2009 Great Recession and the COVID-19 pandemic in 2020. Trends in rail and intermodal shipment data are examined in conjunction with economic indicators, focusing on the extent of drop and recovery of freight volumes of various commodities and intermodal shipments, and the lead/lag time with respect to economic drivers. While impacts from and the rebound from the Great Recessions were slow to develop, COVID-19 produced both profound disruptions in the freight market and rapid rebound, with important variations across commodity types.

Energy-related commodities (*i.e.*, coal, petroleum, and fracking sand), dropped during the pandemic while demand for other commodities (*i.e.*, grain products and lumber, and intermodal freight). rebounded rapidly and in some cases grew. Overall rail freight experienced a rapid rebound following the precipitous drop in traffic in March and April 2020, achieving a near-full recovery in five months. As the recovery proceeded through 2020, intermodal flow, containers moving by rail for their longest overland trips, rebounded strongly, some exceeding 2019 levels. In contrast, rail flows during the Great Recession changed slowly with the onset and recovery, extending over multiple years. Pandemic response reflected the impacts of quick shutdowns and a rapid shift in consumer purchasing patterns. Results for the pandemic illustrate the resilience of U.S. rail freight industry and the multifaceted role it plays in the overall logistics system. Amid a challenging logistical environment, freight rail kept goods moving when other methods of transport were constrained.

**INTRODUCTION**

COVID-19 brought about rapid and complex changes in the demand for freight transportation – decreases, increases, and shifts in products to be moved. Railroads, like all of the transportation industry, are at the mercy of such large-scale economic patterns, as illustrated by past events, most recently the Great Recession of 2007-2009. Positioned in the middle of the economy, the U.S. industry has responded, adapted, and maintained its central role in the most efficient freight and logistics system in the world.

The impacts of the pandemic came as a result of the shutdown of the U.S. economy, both by necessity and by rule, as people stayed home, certain businesses were forced to close while others struggled to ramp up production, employment cratered in some occupations, and thus income dropped, affecting both the scale and scope of consumer purchasing. Many people were buying less; others were buying more and different things, partially stimulated by government emergency subsidies, and making purchases in different ways (e-commerce). At about the same time, U.S. exports and imports were affected by international trade conflicts and tariffs. The freight industry, and the railroads, were greatly affected by these economic forces.

COVID-19 changed the product mix in the logistics system, boosting consumer goods and changing the ways in which they were purchased. While continuing to move most bulk commodities, the railroads had the capacity and flexibility to respond to the rapid expansion of demand for intermodal freight (*i.e.*, consumer goods) in the middle of 2020.

In February, 2022, the U.S. Department of Transportation published a report on supply chain performance under COVID-19, highlighting the dearth of research on supply chain disruptions (*1*). This paper is one of a series examining COVID-affected supply chain performance focusing on resilience of rail intermodal freight industry, including discussion of the impacts of e-commerce (*2*). It provides a high-level analysis of major rail markets during two recent major economic disruptions, comparing the dynamics of onset and rebound of both shipping demand and rail freight supply. It provides insights into the resilience of rail freight that may inform future decisions about freight operations and policies. It uses freight data to illustrate the role of rail freight in responding demand surges, and it suggests opportunities for researchers to pursue studies to analyze and forecast rail freight performance during disruptions. The resulting deeper understanding of the existing and potential role of rail will contribute to advancing the resilience of the nation's supply chain ahead of future disruptions.

The main questions addressed in this paper are these:
1. How did the rail industry fare during two different economic downturns and corresponding recoveries?
2. Specifically, how did the COVID-19 pandemic impact the rail industry, particularly in terms of trade and shipping patterns, as well as changing markets for energy commodities?

Our approach to this effort comprised graphical and statistical analysis of flows, performance, and pricing of rail and truck shipments for both the 2008 Great Recession and COVID-19. We present and analyze the quantitative trends in rail and truck movements and the economic drivers that have occurred since the onset of the COVID-19 pandemic, comparing them to experiences during the Great Recession.

**DATA AND METHODOLOGY**

To explore the similarities and differences between the previous economic recession, known as the Great Recession (December, 2007 – June, 2009) (*3*), and the economic disruption brought by the COVID-19 pandemic, the economic indicators in **Table 1** were selected to explore various rail freight markets. Data sources include the Association of American Railroads (AAR), Freightwaves SONAR, The Federal Reserve Bank, The U.S. Department of Transportation Bureau of Transportation Statistics, the Bureau of Economic Analysis, the Energy Information Administration, and the Census Bureau. The data frequency adopted is monthly.

To study the trends of the rail freight and economic indicators of different types and times, the following parameters were examined:

1. The lag or lead time of the rail freight drop compared with the corresponding economic indicator;
2. The extent of the drop in rail freight in the Great Recession and under COVID-19; and
3. The duration taken for half and full recovery for rail freight volumes, in comparison to the economic indicator.

The rail freight traffic data (*4*) used were the 4-week moving average U.S.-originated traffic excluding the U.S. operations of Canadian and Mexican railroads unless otherwise specified. Periods of U.S. recessions are shaded in charts.

Analysis of total carload and intermodal rail freight is first presented, followed by intermodal freight and other commodities separately.

**Table 1 - Rail Freight Components, Related Trucking Metrics, and Economic Indicators**

| Rail Freight Components | Economic Indicators / [Truck Freight Metrics] | Description |
|---|---|---|
| Overall | Industrial Production *(Federal Reserve)* (*5*) | Real output for U.S. facilities manufacturing, mining, and electric, and gas utilities |
| Intermodal | [Truck Freight Metrics] Truck Tonnage Index *(Bureau of Transp. Statistics)* (*6*) | A relative measure of the total tonnage transported by the trucking industry |
| | [Truck Freight Metrics] Longhaul Outbound Tender Volume Index (*7*) *(Freightwaves SONAR)* | An index of electronically tendered freight volumes over 800 mile length of haul |
| | [Truck Freight Metrics] Longhaul Outbound Tender Reject Index (*8*) *(Freightwaves SONAR)* | A percentage of loads rejected to total loads tendered over 800 mile length of haul, a reflection of the inability or unwillingness of the long haul trucking industry to respond to shipping demands |
| | Real Personal Consumption Expenditures (PCE) - Durable Goods *(Bureau of Economic Analysis)* (*9*) | The value of goods purchased with an average useful life of at least 3 years |
| | Retail Sales: Retail and Food Services / Nonstore Retailers *(U.S. Census Bureau)* (*10, 11*) | The value of retail sales from the Monthly Retail Trade Survey |

| Rail Freight Components | Economic Indicators / [Truck Freight Metrics] | Description |
|---|---|---|
| | Total Business Inventories (U.S. Census Bureau) (*12*) | The value of the end-of-month stocks regardless of stage of fabrication |
| Coal | Coal Production Estimate *(Energy Information Administration)* (*13*) | Estimated total amount of weekly coal production in the U.S. |
| | Industrial Production *(Federal Reserve)* | Real output for U.S. manufacturing, mining, and electric, and gas utilities providers |
| Petroleum Products | EIA Field Production of Crude Oil *(Energy Information Administration)* (*14*) | Estimated volume of crude oil produced during given periods of time in the U.S. |
| | Crude Oil Prices: West Texas Intermediate (WTI) *(Energy Information Administration)* (*15*) | Spot price (dollar per barrel) of a crude stream produced in Texas and southern Oklahoma as reference for pricing of crude streams traded in the domestic spot market at Cushing, Oklahoma |
| Motor Vehicles and Equipment | Domestic Auto Production *(Bureau of Economic Analysis)* (*16*) | Total units of autos assembled in the U.S. |
| Lumber and Wood Products | Housing Starts: Total: New Privately Owned Housing Units Started *(Census Bureau)* (*17*) | Number of privately owned housing units start which occur when excavation begins for the footings or foundation of a building |

## HISTORICAL CONTEXT FOR RAIL SERVICE RESILIENCE UNDER THE GREAT RECESSION

This section presents the patterns of rail freight volume in relation to the general economy during the Great Recession. While this depicts its general development, the insights such as factors for the drop/recovery pattern are further discussed in the next section in comparison with the COVID-19 pandemic.

### Total Carload and Intermodal Rail Freight

Industrial Production (IP) was chosen as the economic indicator to evaluate the response of total rail freight, including both carload and intermodal shipments, at the macroscopic level. Selected commodities and intermodal rail freight were studied in a similar manner with respective economic indicators.

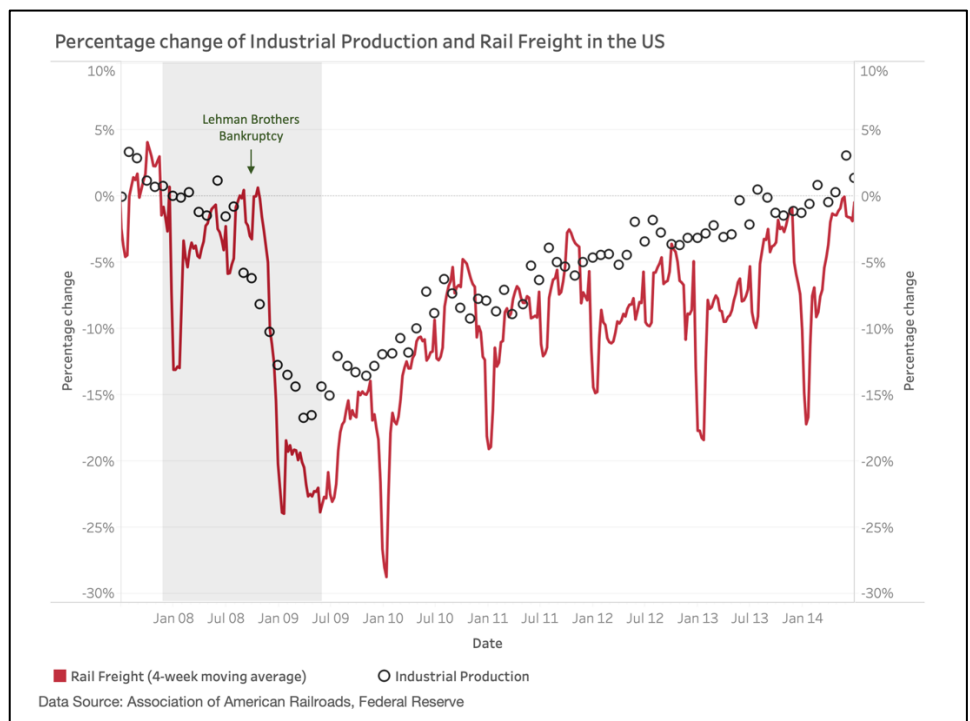

**Figure 1 – Percentage change of Industrial Production and Rail Freight in the US (2007-2014)**

As shown in **Figure 1**, the drop in IP started with the onset of the recession in December, 2007, and gained pace in August, 2008. Meanwhile, the total rail freight in 2008 stayed around 3% below the start of the year and then decreased significantly in October, 2008, which lagged behind IP by two months. The levels of decrease were 17% for IP and 23% for rail freight. Although rail freight reached its bottom level in January, 2009, which coincided with the usual trough at the end of the year, recovery did not start before IP also reached its bottom in May, 2009, *i.e.*, rail freight recovery lagged industrial production by about 5 months.

The half recovery of rail freight took a year, the same time as that of IP. It took rail freight 5 years to achieve full recovery, one year later than IP.

**Figure 2** illustrates the proportions of rail freight by commodity type before, during, and after the recession. Intermodal and coal constituted around 70% of rail freight by total carload and intermodal units.

**Intermodal (IM) Rail Freight**

IM rail freight has grown by 70% over the past 20 years. Real Personal Consumption Expenditure (PCE) - Durable Goods was chosen to reflect the changes in the demand side for IM freight. This is because durable goods, excluding daily consumption items such as groceries and gas, are more sensitive to economic changes.

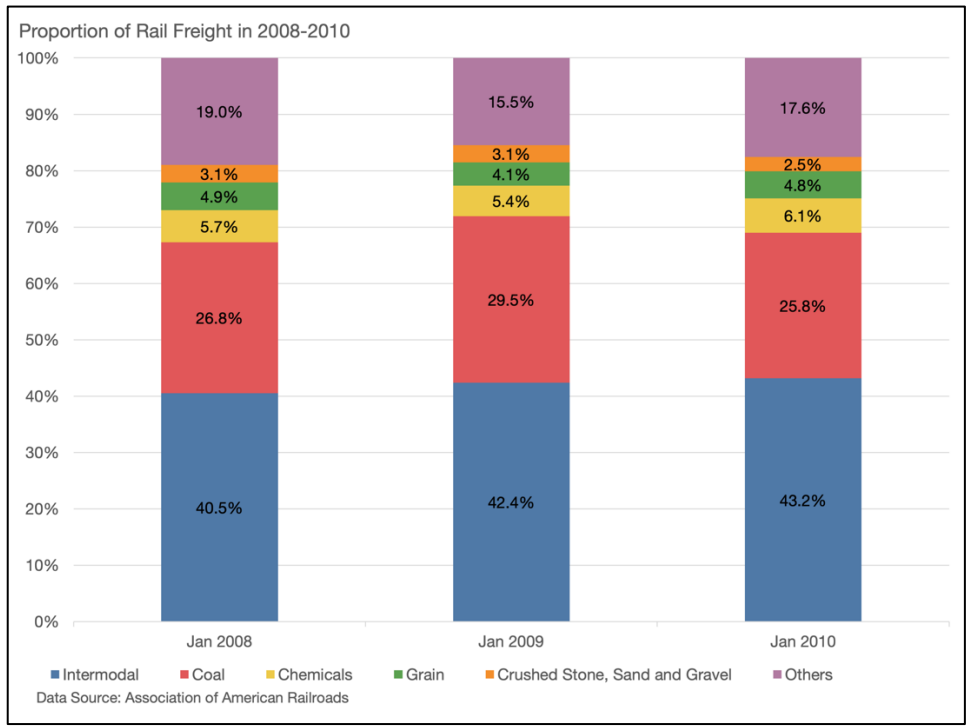

**Figure 2 - Proportion of Rail Freight by Commodity in the US (carloads/intermodal units 2008-2010)**

**Figure 3** shows that as the financial crisis unfolded from August to October, 2008, Real PCE – Durable Goods gradually decreased by 13%. The IM freight volume fell by 26% within three months from October, 2008 to January, 2009, which lagged behind the PCE for four months and coincided with the usual year-end trough. This lag is likely due to products still being in the supply pipeline even after production slows.

Recovery for IM freight began in March, 2009, after the durable goods expenditures hit bottom in the same month. This was similar to the trends in total rail freight and IP. Half of the losses were recovered in 10 months for Real PCE – Durable Goods and 12 months for IM. The full recovery of IM was completed in 2.5 years, which was faster than the 5 years of overall freight and at the same time when the durable good expenditure returned to the pre-recession level.

**Figure 4** compares the same economic indicator with truck freight volume as captured by the Truck Tonnage Index from the Bureau of Transportation Statistics. As a competing mode, truck freight tracked closely with IM, albeit with a smaller drop of 14%, similar to that in Real PCE - Durable Goods. In contrast to IM, which lagged behind the PCE indicator by four months, the truck response was rapid, showing no lag for the Truck Tonnage Index to fall and reach the bottom. This suggests, not surprisingly, that truck operations and markets were able to respond more rapidly than rail to demand shifts.

Nevertheless, the trucking recovery time was analogous to IM, with minimal lag or lead to Real PCE – Durable Goods, taking 10 months for half recovery and 2.5 years for a full recovery. This was different from the initial drop potentially because there was ample time for trucking firms to match the slow recovery of demand.

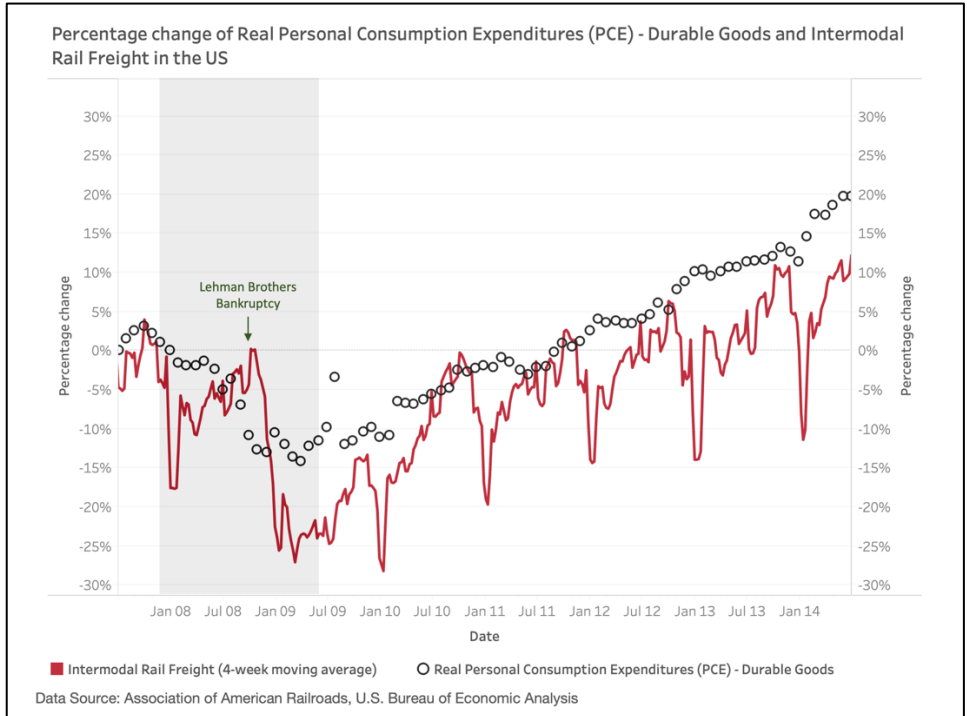

**Figure 3 – Percentage change of Real Personal Consumption Expenditure (PCE) – Durable Goods and Intermodal Rail Freight in the US (2007-2014)**

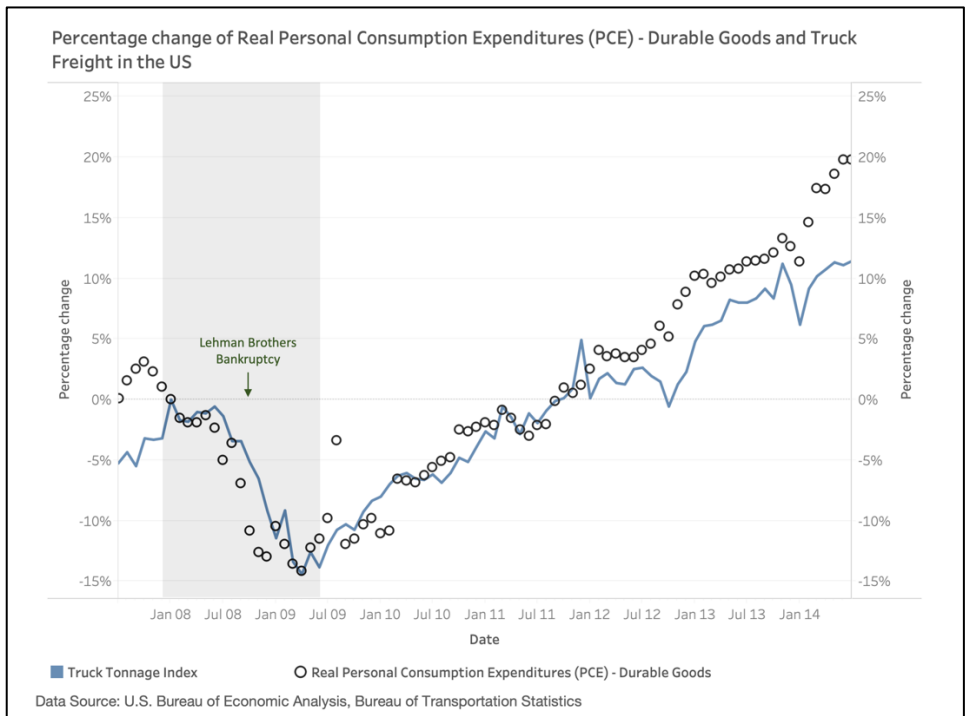

**Figure 4 – Percentage change of Real Personal Consumption Expenditure (PCE) – Durable Goods and Truck Freight in the US (2007-2014)**

**Figure 5** shows trends in business inventories, which started dropping at around the peak of the recession, and then fell around 15% over two years, reflecting retailers' expectation of reduced needs for inventories due to lower sales. The trough in inventories lagged behind the IM freight for around half a year as shipments slowly refilled inventories. The full recovery of inventories took around two years.

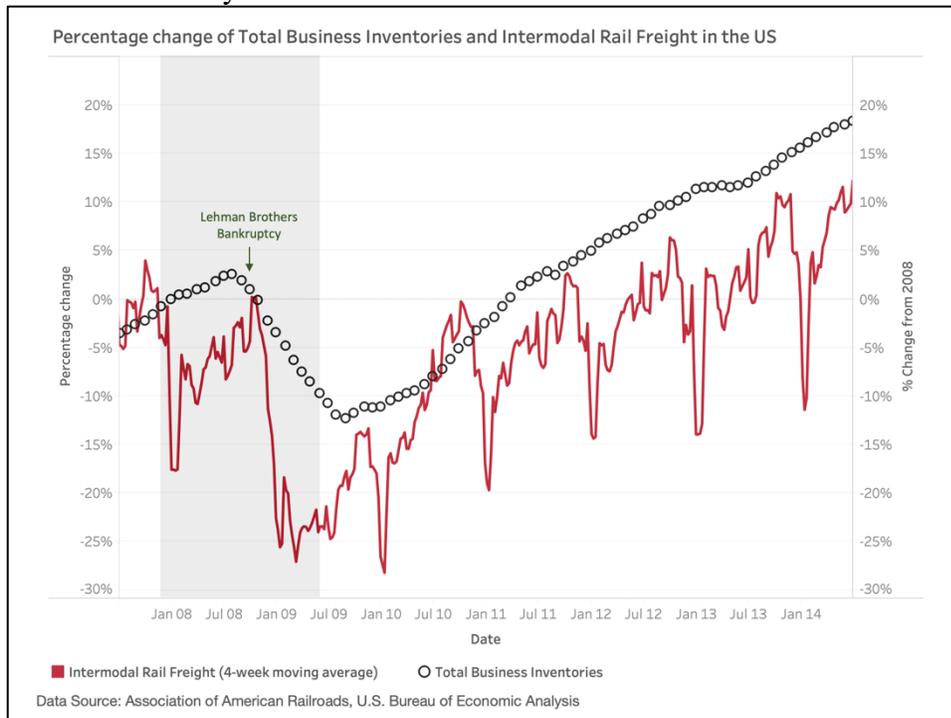

**Figure 5 - Percentage change of Total Business Inventories and Intermodal Rail Freight in the US (2007-2014)**

## Coal Rail Freight

The movement of coal by rail closely follows that of coal production, as rail is the major mode for moving coal. Coal rail freight remained steady in the first part of the recession until production peaked in late 2008 (**Figure 6),** followed by a long-term decline of 45% between 2009 and 2020, during which period coal production fell by 35% between 2009 and 2020. This long-term decrease starting in the previous decade, on average 4% yearly, can be explained by competition from natural gas followed by increasing production of oil in the U.S.. Coal production and shipping are closely linked to the price of competing energy sources for electricity generation. The crude oil price (WTI) fell 70% in the second half of 2008, reaching its bottom in January, 2009, which led coal production and rail freight by four months. Coal production and rail freight dipped further in 2016 when major coal producers filed for bankruptcies following the worldwide drop in energy prices.

Industrial Production (IP), another indicator shown in **Figure 7**, illustrates the fluctuation of coal rail freight in response to aggregate industrial activities during the recession. As compared with the coal freight, the major drop in IP started 3 months earlier and reached the bottom 2 months earlier with a decrease of 17%.

While IP recovered gradually until 2014, coal rail freight showed a general downward trend as driven by the market factors discussed above. In summary, coal moved by rail during the Great

Recession was affected more by the long term market demand for coal against the competition than by the condition of the economy.

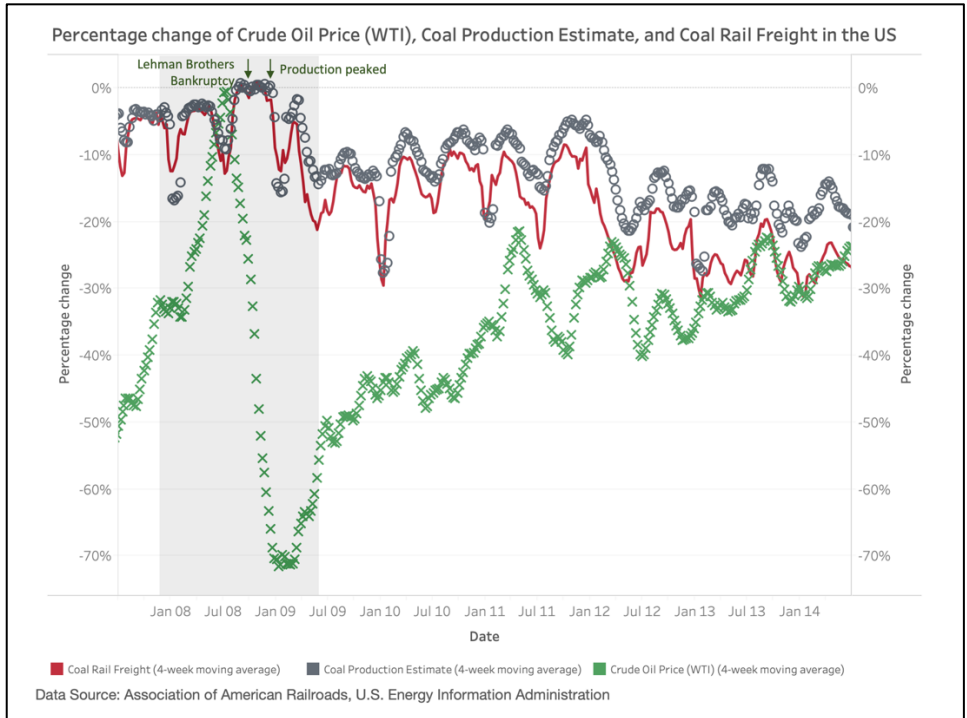

**Figure 6 – Percentage change of Crude Oil Price, Coal Production Estimate and Coal Rail Freight in the US (2007-2014)**

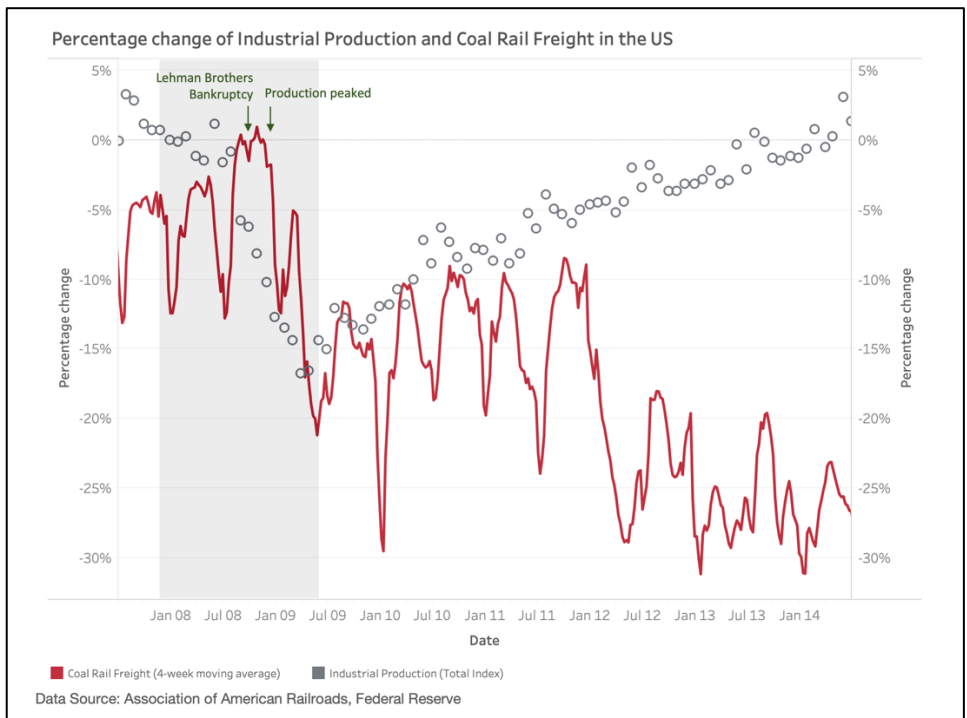

**Figure 7 – Percentage change of Industrial Production and Coal Rail Freight in the US (2007-2014)**

## Petroleum Rail Freight

The petroleum component of rail freight depends mainly on the domestic production of crude oil, with pipeline as the transportation competitor, which in turn was influenced more directly by fluctuations in the energy market and local production than by the economic recession.

In **Figure 8**, with 2008 as the baseline, the volume of petroleum products moved by rail freight increased by 160% until 2014, while crude oil field production grew by 80% in the same timeframe. This increment followed the growth in shale oil production in the U.S. Afterward, rail freight dropped by nearly half until 2017, when oil production contracted by 10% and some traffic shifted to newly-built pipelines.

During the worldwide energy price crises in 2008 and 2014 as signaled by the sharp fall in the crude oil price (WTI) by more than 50%, crude oil production decreased only temporarily and moderately (less than 20%) due to the less elastic nature of production in a short term, which involved fixed investment in oil pumps and fracking facilities. This was in contrast to the larger decrease in the rail freight, which may suggest the shifting of freight share between rail and pipeline as the volume fluctuated.

In summary, just as in the case of coal by rail, petroleum products moved by rail seem to have been affected more by global petroleum production and pricing than by the Great Recession.

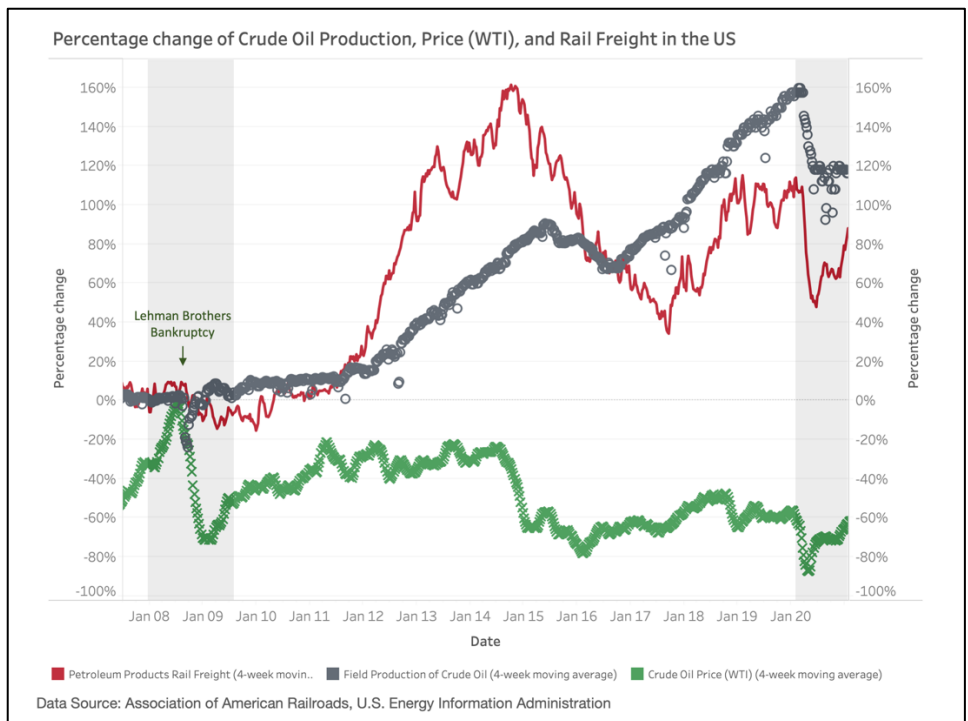

**Figure 8 - Percentage change of Crude Oil Production, Price (WTI), and Rail Freight in the US (2007-2020)**

## Motor Vehicles and Equipment Rail Freight

The automotive industry was among the most impacted in the Great Recession. Soon after the financial crisis culminated with the bankruptcy of Lehman Brothers in 2008, domestic auto production dipped by 70% in three months, as shown in **Figure 9**. Rail freight followed with a decrease of 60%.

After the federal government bailout of the American auto industry in 2009, both automotive production and auto-related rail freight recovered by half in nine months. Full recovery took another 2.5 years, until 2012.

**Figure 10** shows Real PCE – Durable Goods as a demand indicator to reference the trends in auto rail freight, both of which followed a similar timeline of decline and recovery. The PCE dropped by 13% from August to October, 2008, within the same timeframe as the decrease in rail freight by 60%. The half recovery of both took around 10 months, and the full recoveries were completed in 2.5 years.

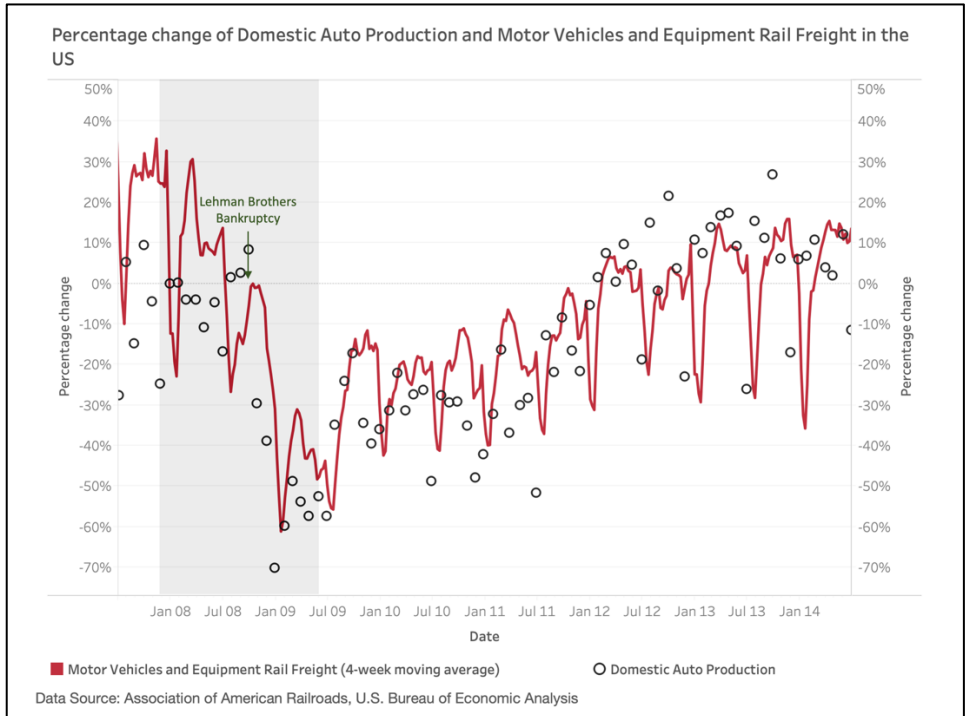

**Figure 9 – Percentage change of Domestic Auto Production and Motor Vehicles and Equipment Rail Freight in the US (2007-2014)**

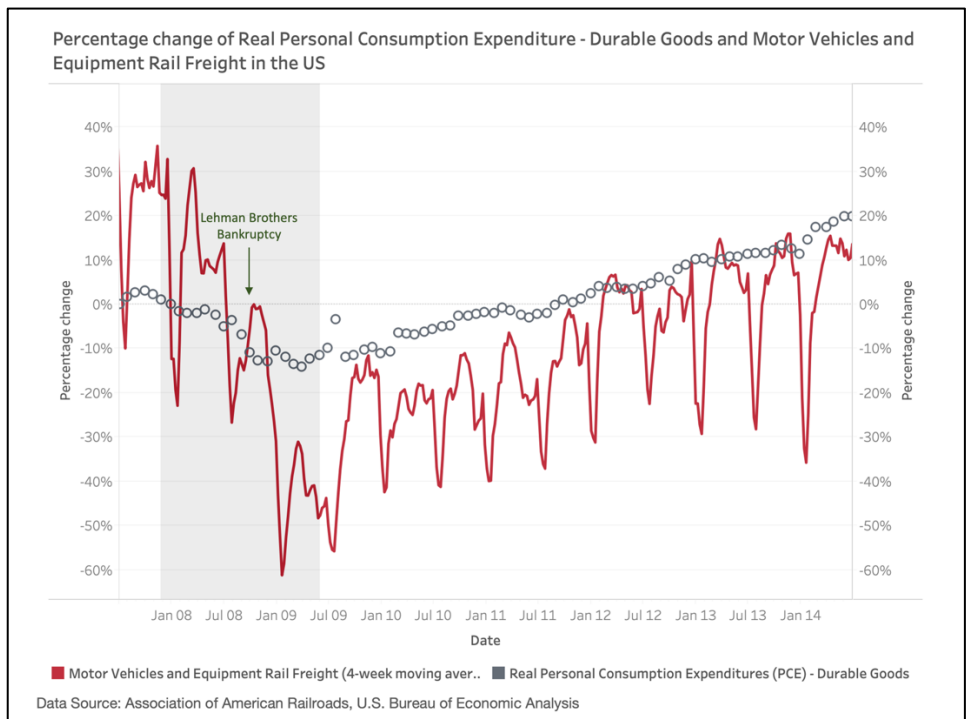

**Figure 10 - Percentage change of Real Personal Consumption Expenditure (PCE) – Durable Goods and Motor Vehicles and Equipment Rail Freight in the US (2007-2014)**

## Lumber and Wood Products

Wood is a major construction material for single-family homes, making housing starts for new, privately owned units a good indicator to reference variation in lumber and wood products moved by rail. As shown in **Figure 11**, following the subprime mortgage crisis starting from 2007, housing starts fell by more than 35% from September to December, 2008. The drop in lumber and wood products moving by rail occurred in the same time frame and amounted to more than 40%.

After reaching the bottom, housing starts lingered at a low level for 2.5 years and then took two more years for full recovery. Rail freight showed a similar trend and fully recovered in 4.5 years.

The analyses of the movement of energy products and building materials by rail shows that the rail freight response to the Great Recession logically followed the relevant driving variables in both the economic decline and recovery, as well as long term trends in fuel sourcing and pricing. Notable in the patterns was the substantial lag in the response of rail volumes compared with the economic drivers, reflecting the effects of inventories both in motion and in warehouses, as well as the multi-year period required for recovery of both the economic indicators and rail shipments to pre-recession levels.

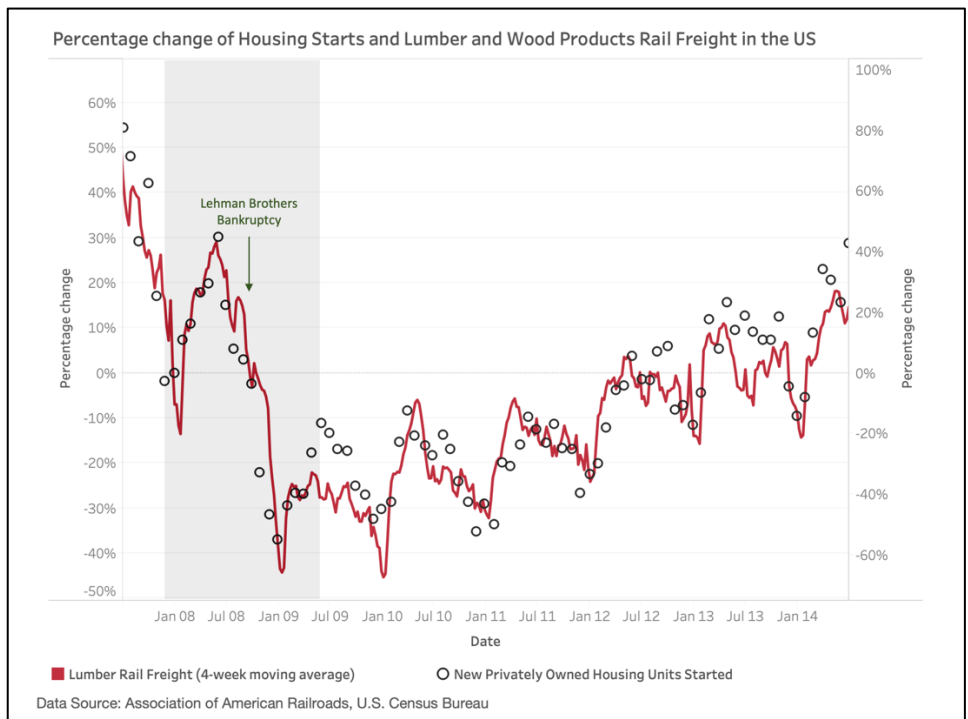

**Figure 11 - Percentage change of Housing Starts and Lumber and Wood Products Rail Freight in the US (2007-2014)**

## THE RECENT DOWNTURN: IMPACT OF COVID-19 PANDEMIC ON RAIL AND INTERMODAL

In this phase of the analysis, the patterns of disruption due to COVID-19 were examined in search of similarities and differences with the Great Recession. As before, selected economic indicators were utilized to relate the response of rail freight to the disruption of specific economic factors, first as a whole for all carload and intermodal freight, followed by breakdowns into intermodal and selected commodity flows.

### Total Carload and Intermodal Rail Freight

As shown in **Figure 12**, rail freight suffered a significant drop of 18% in the early phase of COVID-19 lockdown from February to April, 2020, closely following the drop in industrial production (IP) of 18% from February to March. This illustrated the short-term disruptive effect of the COVID-19 lockdowns in contrast to the two-month lag of rail freight at the onset of the economic recession in 2008.

As the lockdown was relaxed following the re-opening of various states, the rebounds of the rail freight and IP were rapid, recovering half of the losses within three months. Rail freight grew further to the pre-COVID level, achieving full recovery in five months. Nevertheless, IP was still 3% below the pre-pandemic at the end of 2020. The deviation of total rail freight from IP was supported mainly by the growth in intermodal traffic, driven, in turn, by imports, *i.e.*, products produced offshore.

Before the coronavirus outbreak, there was a 10% decline in rail freight from the peak in 2018 during the China-United States trade war. As of January, 2021, while rail freight recovered all the losses since the pandemic, it had not yet reached the previous peak.

**Figure 13** shows the proportion of rail freight by commodity types before and during the pandemic. The portion of intermodal freight rose continuously from the last recession to more than 56% of the US rail freight market at the start of 2021. By contrast, the coal freight percentage dropped from more than 25% in 2010 to around 12% in 2021.

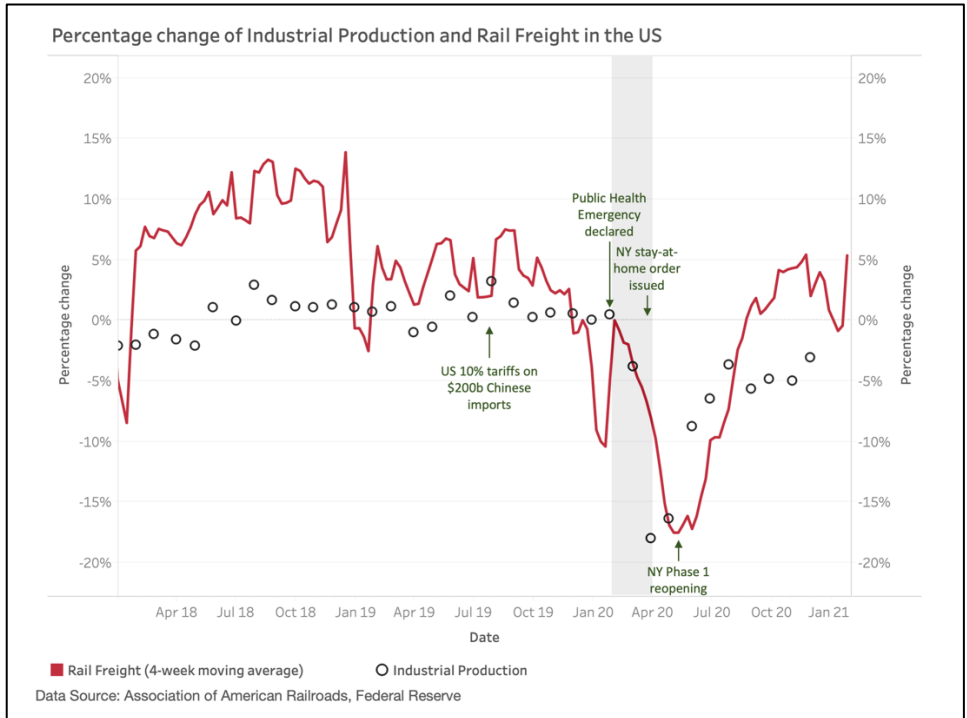

**Figure 12 – Percentage change of Industrial Production and Rail Freight in the US (2018-2021)**

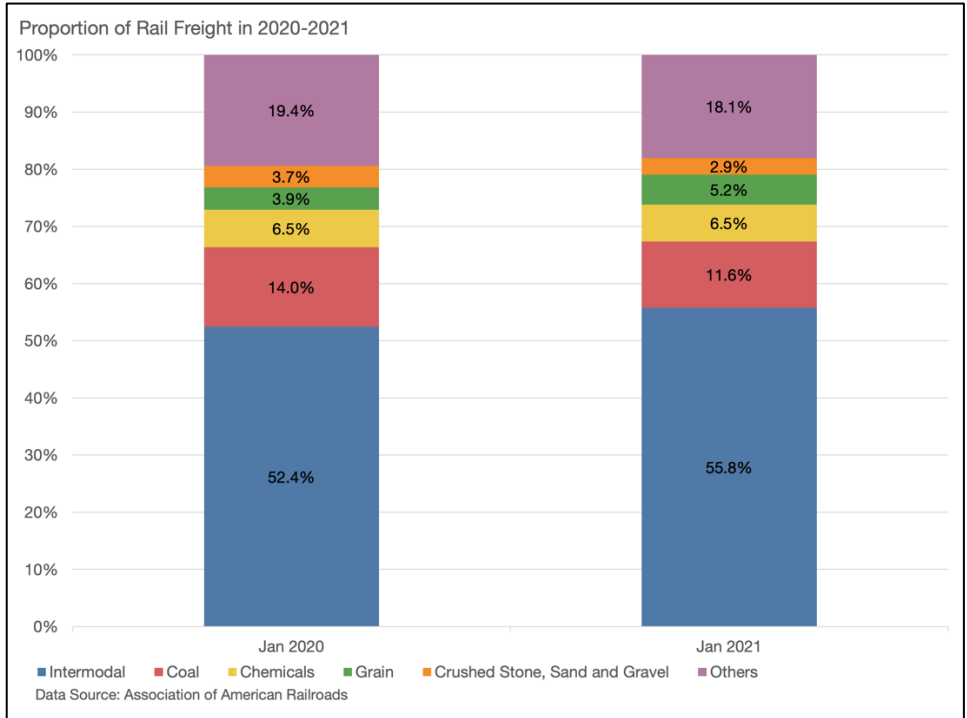

**Figure 13 - Proportion of Rail Freight by Commodity in the US (carloads/intermodal units 2020-2021)**

## Intermodal (IM) Rail Freight and the Pandemic

**Figure 14** shows the response of IM rail freight in the pandemic. During the early stage of the virus outbreak, from February to April, 2020, the IM volume suffered a 17% drop, while Real PCE - Durable Goods decreased by 22%. IM freight showed a V-shaped recovery following full recovery of the durable goods consumption in May. The IM volume returned to the original level in three months (July, 2020), lagging consumption by two months, as inventories were drawn down. In the third quarter of 2020, the freight volume and the PCE indicator increased further to more than 10% higher than the start of the year. This growth was supported by the new demand in e-commerce induced by the pandemic and lockdown measures – preference to avoid exposure in stores and store staffing reductions and closures.

Before the COVID-related events, IM rail freight had shown a gradual decrease of 5% since 2018, which coincided with the China-United States trade war. Within IM, trailer rail freight experienced a strong increase of nearly 40% beyond the initial rebound in 2020, closely following retail sales. Trailer freight provides more direct-to-store transportation, avoiding competition for chasses necessary for moving containers over the road, and themselves in short supply (*2*).

**Figure 15** shows the relationship between total business inventories and IM freight volume. Inventories dropped from the start of the pandemic and reached its trough of 5% fall off in May, 2020. The lag of inventories behind IM freight was logical as IM freight caught up with reduced stocks, not unlike the Great Recession, but at a much faster pace. Under COVID, inventories did not fully recover at the end of 2020, as the boost in demand limited business capacity to replenish stock.

Truck freight (Tonnage Index) experienced a smaller drop of 8% compared with IM (17%) and Real PCE – Durable Goods (22%) by April, 2020, as shown in **Figure 16**.

The half recovery of the Truck Tonnage Index took three months and had not yet achieved full recovery as of December, 2020. It is important to note that the Truck Tonnage Index covers all motor goods transport. During the initial stages of COVID-19, short-distance goods transport was subject to a smaller decrease, making local deliveries of supplies of living essentials (*e.g.*, groceries), which contributed to the relatively mild drop in truck freight in April, 2020.

To focus on long-haul truck freight (longer than 800 miles), which provides more insights into the market competition for IM, the Long Outbound Tender Volume Index and the Rejection Index were extracted from Freightwave SONAR (**Figure 17** and **Figure 18**). The long-haul tender volume decreased by 30% in April, 2020, matching the timing of the drop in the Truck Tonnage Index. From April to October, the Long Outbound Tender Volume Index increased by 100%, echoing the growth of Real PCE – Durable Goods and the shift of expenditures from services to goods during the pandemic.

The Longhaul Outbound Tender Rejection Index indicates the excess of demand over supply for truck freight for longer journeys. Since September, the rejection index increased to more than 25% - more than a quarter of total loads tendered were rejected. This suggests the capacity limitation of truck freight and likely shifted some demand to IM, a capacity benefit reflecting the ability of the railroads to respond to market volatility. Reports indicated that truck capacity was limited because of labor shortages (*2*).

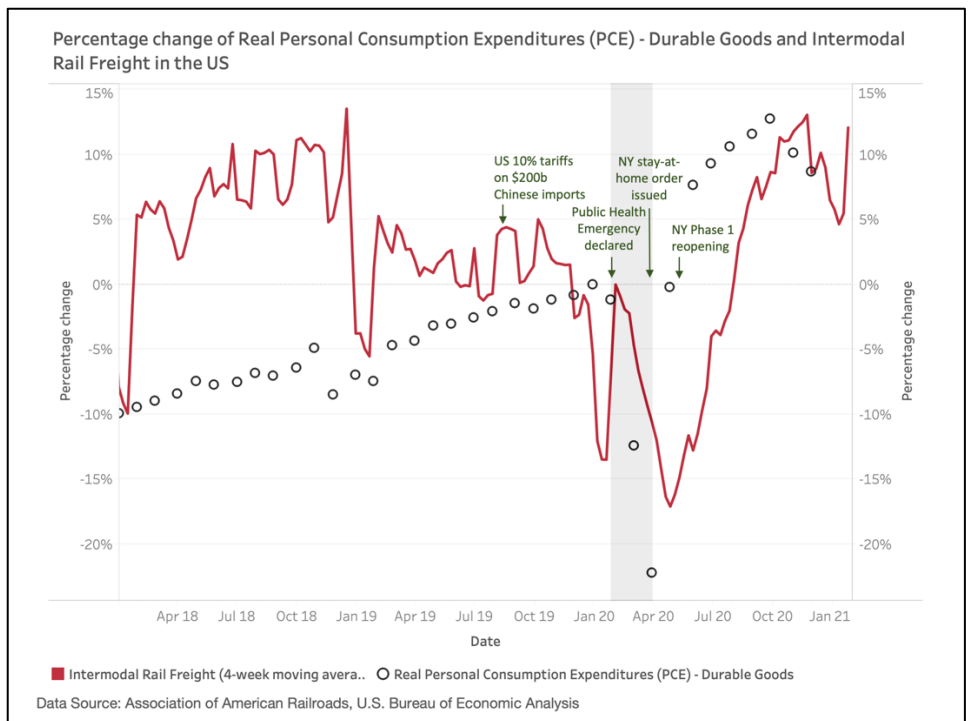

**Figure 14 - Percentage change of Real Personal Consumption Expenditure (PCE) – Durable Goods and Intermodal Rail Freight in the US (2018-2021)**

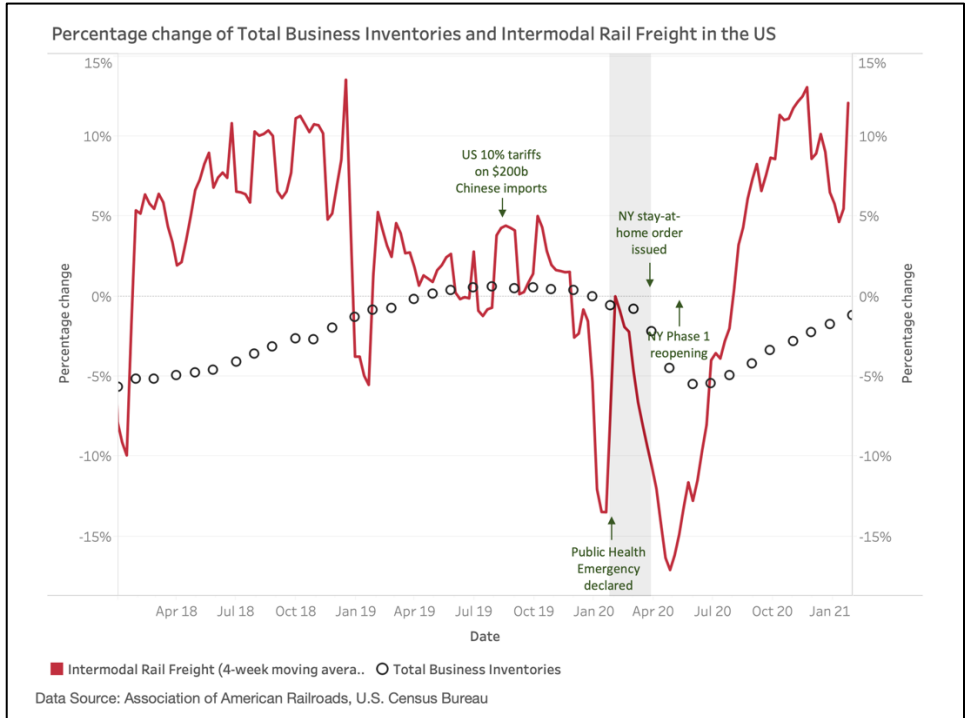

**Figure 15 – Percentage change of Total Business Inventories and Intermodal Rail Freight in the US (2018-2021)**

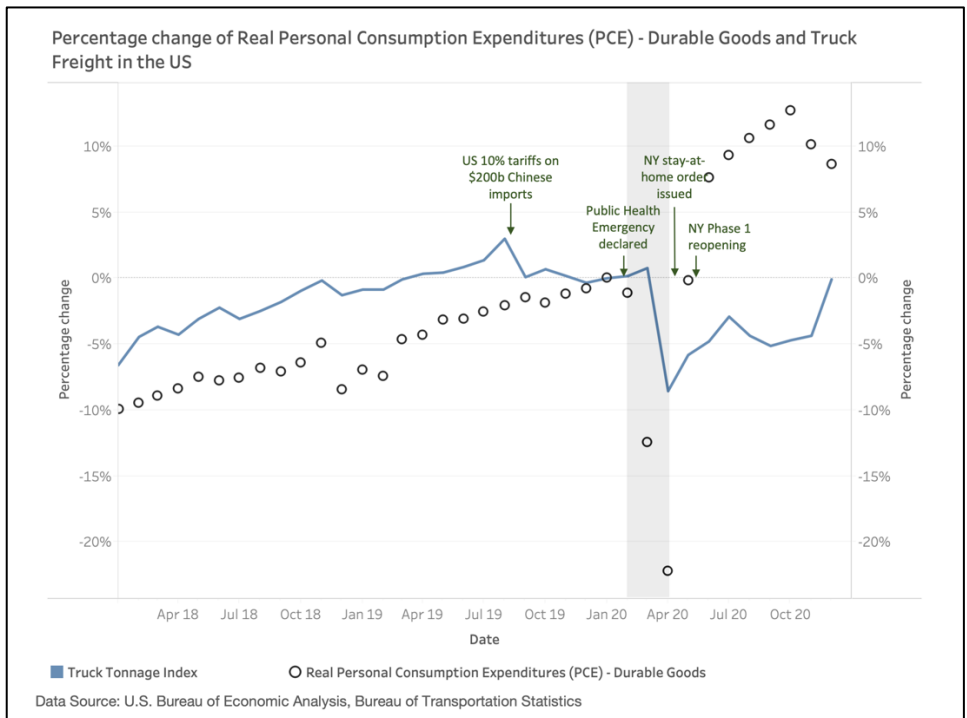

**Figure 16 – Percentage change of Real Personal Consumption Expenditure (PCE) – Durable Goods and Truck Freight in the US (2018-2021)**

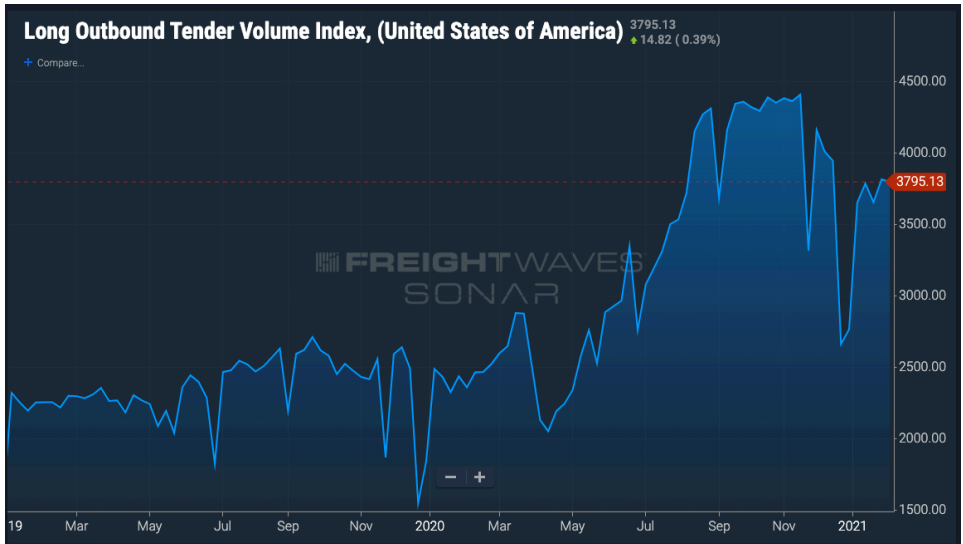

**Figure 17 – Long Outbound Tender Volume Index (USA) (2019-2021)**

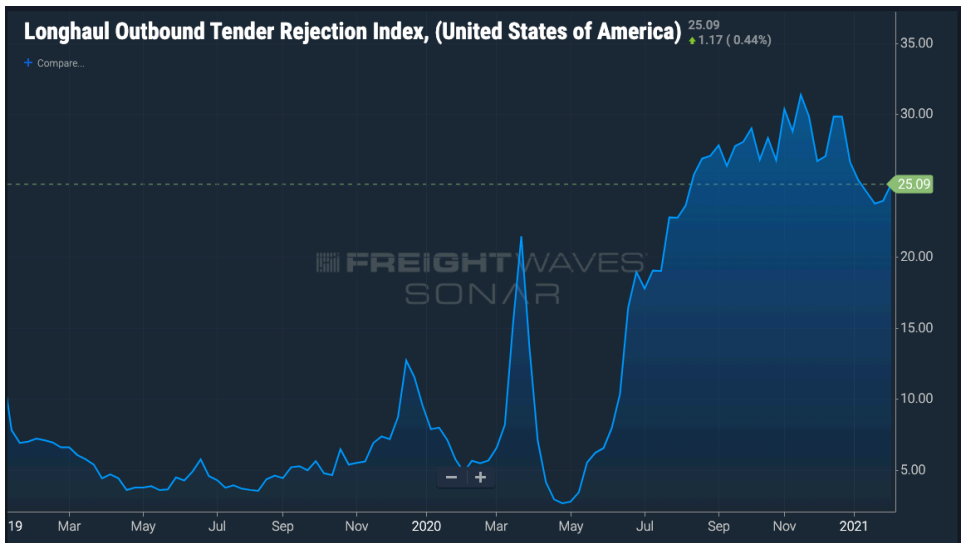

**Figure 18 – Long Outbound Tender Rejection Index (USA) (2019-2021)**

*Intermodal Trailer Rail Freight*

The proportion of trailer rail freight in overall IM had exhibited a generally decreasing trend over the last decade, with container freight accounting for an increasing share of the IM volume. However, Error! Reference source not found. **Figure 19**, reveals that after the initial lockdown-related drop of more than 20% in April, 2020, trailer volume recovered fully by June, 2020 and further increased significantly to a level 40% higher than at the start of 2020.

The change in total retail sales experienced a similar trend, with an initial decrease of 20% in April, 2020 and prompt rebound in June, 2020. The level, however, did not increase further to the same extent as trailer freight did. In contrast, the non-store components of retail sales in Error! Reference source not found. showed a sustained increase of 30% since May, 2020 compared to the pre-COVID level. This supports the previous inference that e-commerce was a major factor fueling IM growth. The press to move products to online shoppers quickly may have motivated cargo owners to transload import shipments from containers to trailers at or near ports, amplified by pressure from ocean carriers to hold scarce containers near ports for quick return to origins.

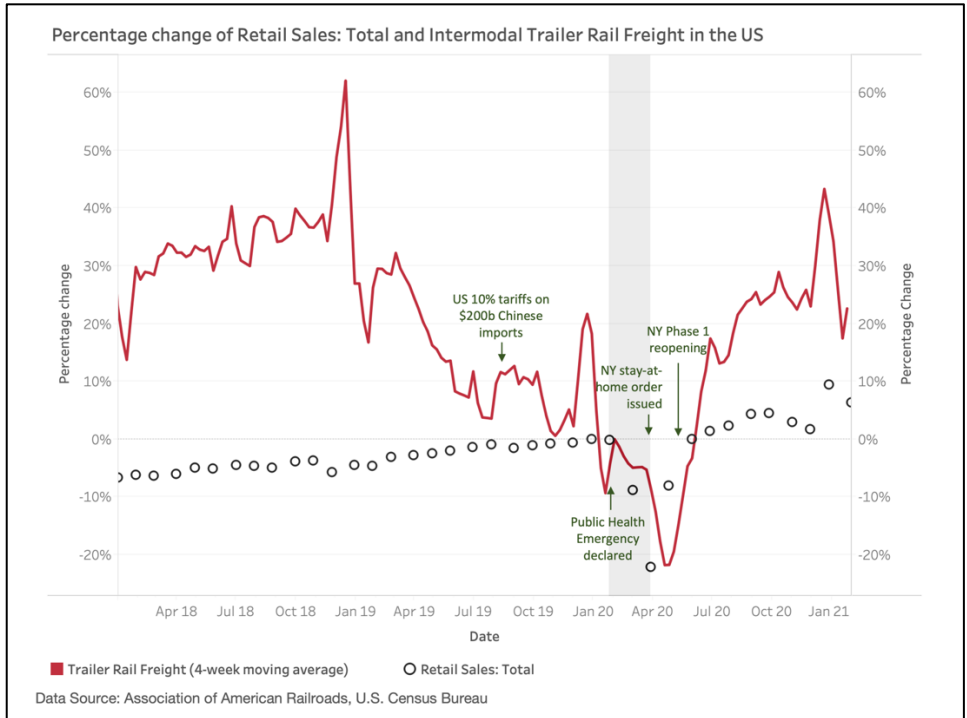

**Figure 19 - Percentage change of Retail Sales: Total and Intermodal Trailer Rail Freight in the US (2018-2021)**

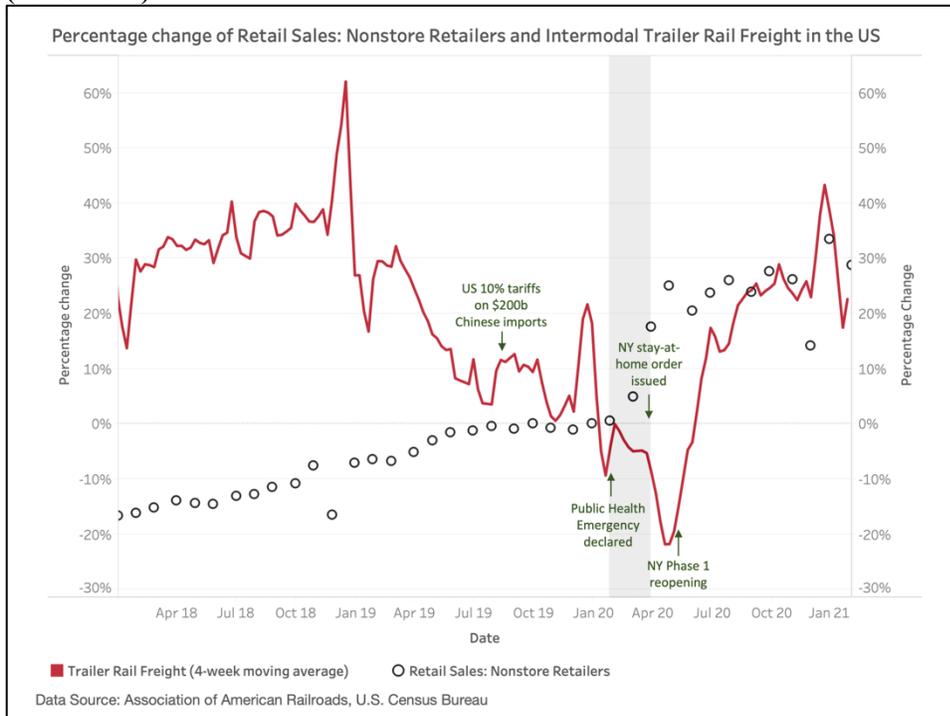

**Figure 20 - Percentage change of Retail Sales: Nonstore Retailers and Intermodal Trailer Rail Freight in the US (2018-2021)**

In summary, the rapid onset of lockdowns led to rapid sales reductions and drop in IM freight; but quick rebounds in sales and volume were also observed as people shifted and accelerated purchasing. IM freight lagged behind inventories drop-off but led rebuilding of stocks. While there were rapid demand pressures on both rail and trucking, under the quick changes in the economic drivers (COVID vs recessions), the ability of rail to respond was apparent.

**Coal Rail Freight**

**Figure 21** shows that, following the 45% drop before the year 2020 mentioned earlier, rail shipments of coal, together with the Coal Production Estimate, dipped by more than 30% during the pandemic and the coincident worldwide fall in energy prices. The crude oil price (WTI) began falling in January, 2020 and turned negative briefly in April, which led the trough of coal production and freight by one month. Subsequently, coal production and rail freight showed a V-shape rebound, recovering half of the losses in three months but remaining 10% below the pre-pandemic level.

While the short-term disruption in coal production brought on by COVID-19, likely because of the temporary shutdown of mines and labor restrictions, was recovered, the long-term decline in coal production and coal freight by rail persisted.

Industrial Production, shown in **Figure 22**, dropped rapidly from February to April, 2020 by nearly 20%, reaching the bottom one month earlier than the coal rail freight. This shorter lead time, attributed to the quick disruption of the initial lockdown in the pandemic, contrasted with the slower response during the Great Recession. IP recovered half of the losses two months later in June, 2020, leading rail freight also by two months.

**Figure 21 - Percentage change of Crude Oil Price (WTI), Coal Production Estimate, and Coal Rail Freight in the US (2018-2021)**

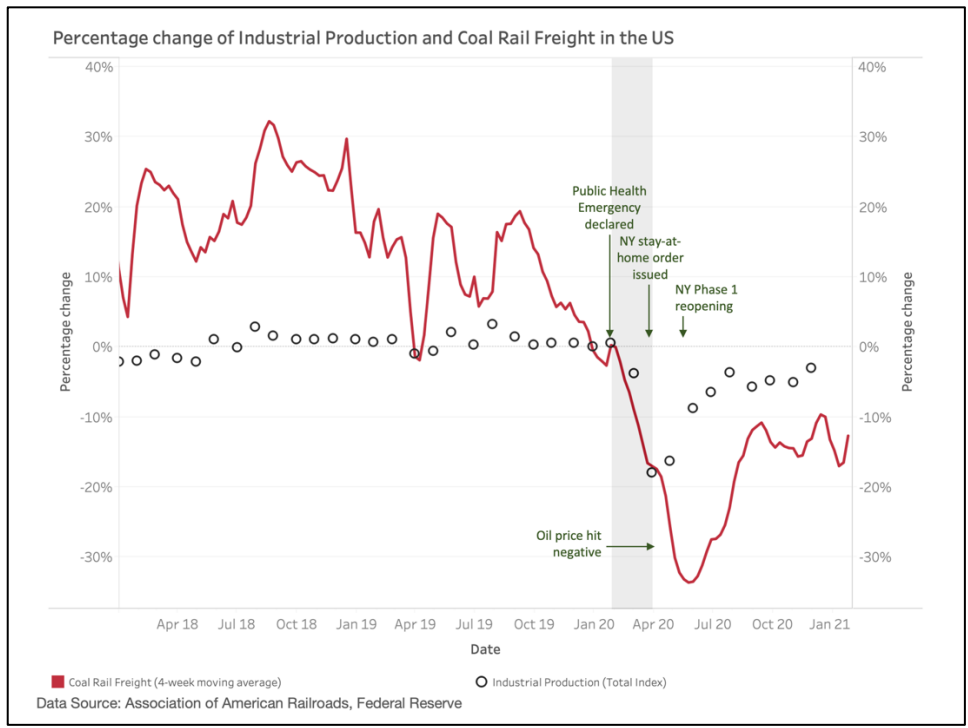

**Figure 22 - Percentage change of Industrial Production and Coal Rail Freight in the US (2018-2021)**

## Crude Oil Production

**Figure 23** shows that the petroleum products moved by rail dropped by 25% from March to May, 2020, following the decrease of crude oil production by 15% in the same period. The WTI oil price started dropping in January and reached bottom and briefly going negative in April, leading rail freight by a month.

While crude oil rail freight recovered by a small amount to return to 10% below the pre-COVID level, oil production remained around 15% lower than the start of 2020. The reasons included reduced energy demand in the U.S., as industrial production had not yet fully recovered. Low energy prices also prompted the shutdown of some drill rigs.

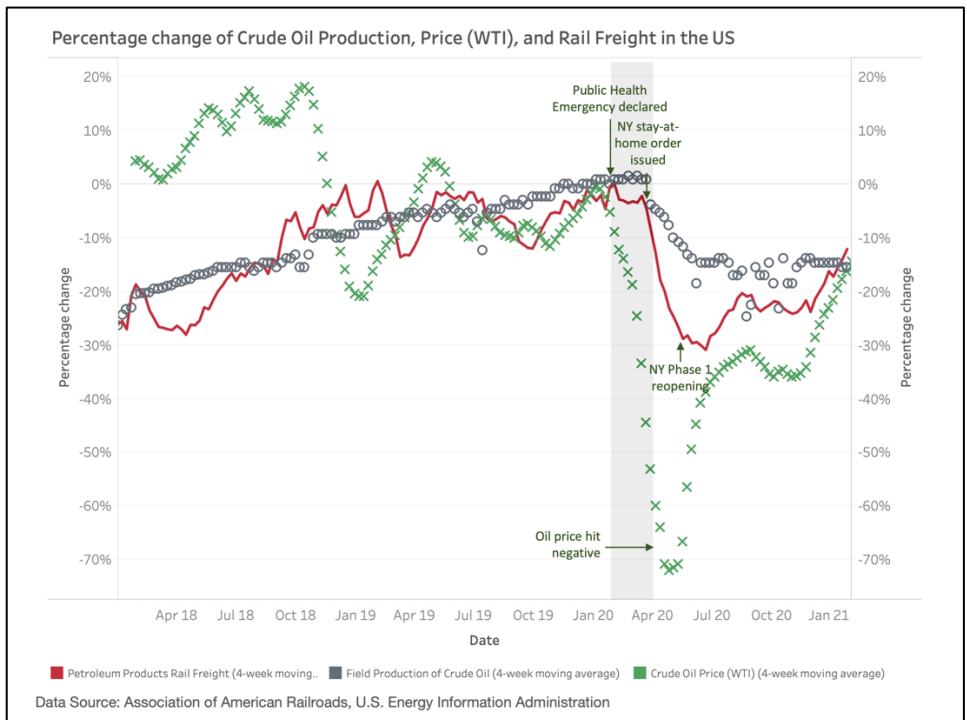

**Figure 23 - Percentage change of Crude Oil Production, Price (WTI), and Rail Freight in the US (2018-2021)**

## Motor Vehicles and Equipment Rail Freight

**Figure 24** shows the sharp V-shaped trend for both Domestic Auto Production and Motor Vehicles and Equipment Rail Freight in 2020. Similar to the previous recession, freight volume closely followed the movement in production with a lag of around one month.

In the first quarter of 2020, the imposed lockdown reduced vehicle production by 99%, while rail freight fell by more than 80%. As reopening spread across the nation, both production and rail freight rose to the original level within three months. However, since August, the production measure remained 5% below the previous year as of December, 2020, which might be related to the constraints in supply chain due to chip shortage.

Real PCE – Durable Goods led the drop of automobile rail freight by one month (**Figure 25**), with a bottom at 20% below the pre-pandemic level in April, 2020. The full recovery was rapid and took only one month, while the rail freight recovered three months later. This highlighted a difference compared with the Great Recession, in that the COVID-19 downturn was driven mainly by a temporary production shock, rather than by a drop in demand.

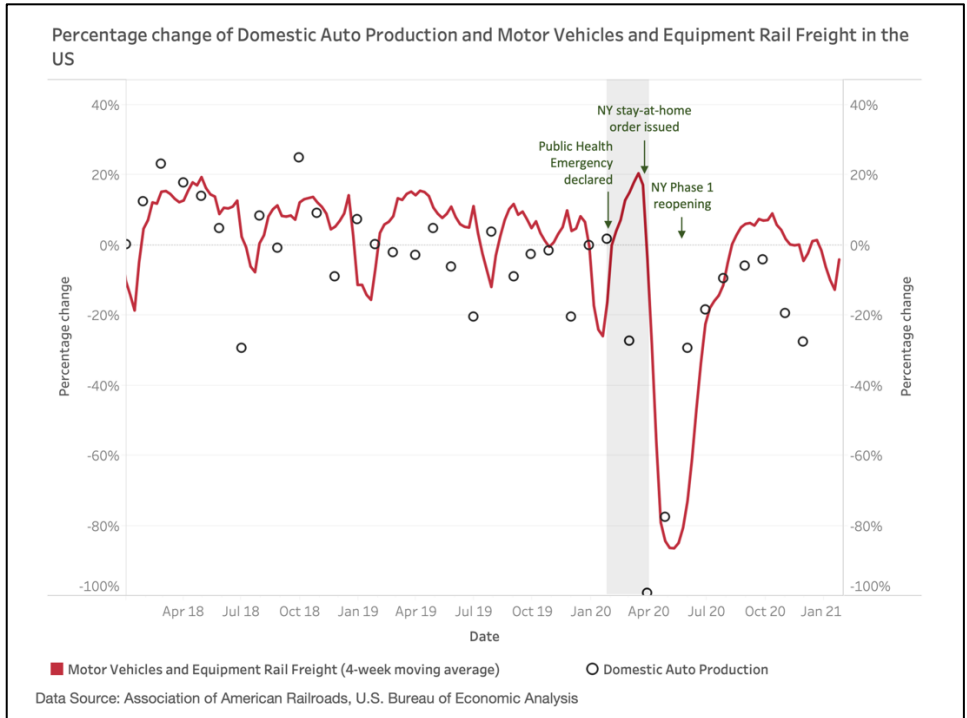

**Figure 24 - Percentage change of Domestic Auto Production and Motor Vehicles and Equipment Rail Freight in the US (2018-2021)**

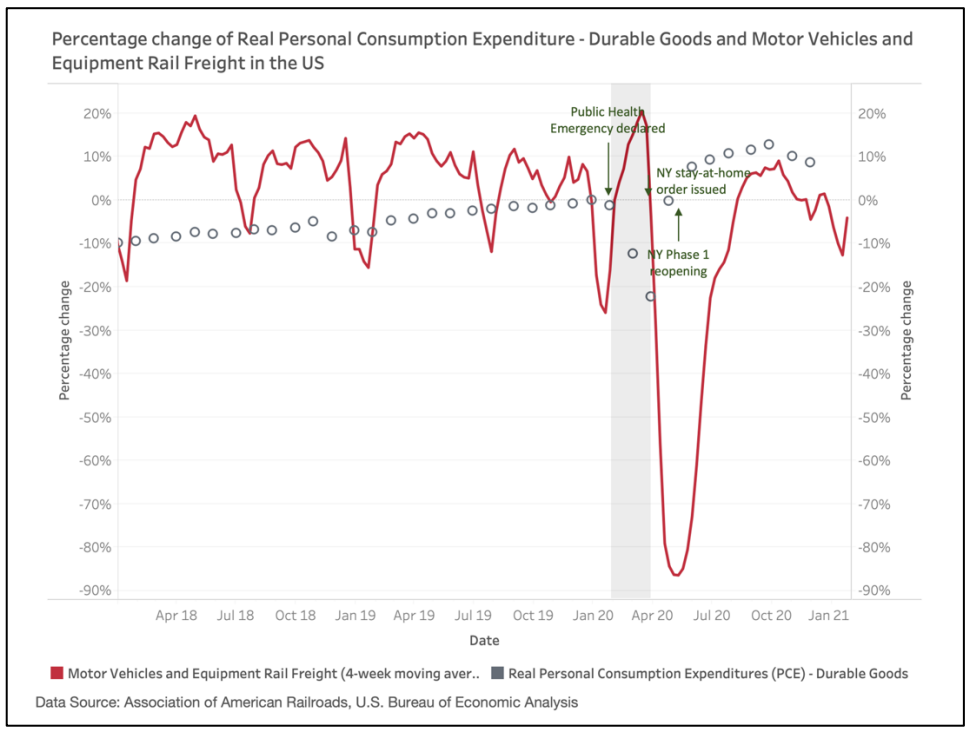

**Figure 25 - Percentage change of Real Personal Consumption Expenditure (PCE) – Durable Goods and Motor Vehicles and Equipment Rail Freight in the US (2018-2021)**

**Lumber and Wood Products**

From January to April, 2020, the number of new private housing starts fell by 25% as the pandemic began to emerge (Figure 26). Different from synchronous drop during the recession in 2008, rail freight of lumber and wood products lagged by two months, falling by 15%.

In June, 2020, the ease in lockdown brought housing starts to full recovery in two months. Similarly, rail volume recovered fully two months after the bottom level in August, while starts of new private housing units rose higher than the pre-pandemic level, compensating for the disruption during the lockdown. As of December, 2020, housing starts remained at their highest level since 2008.

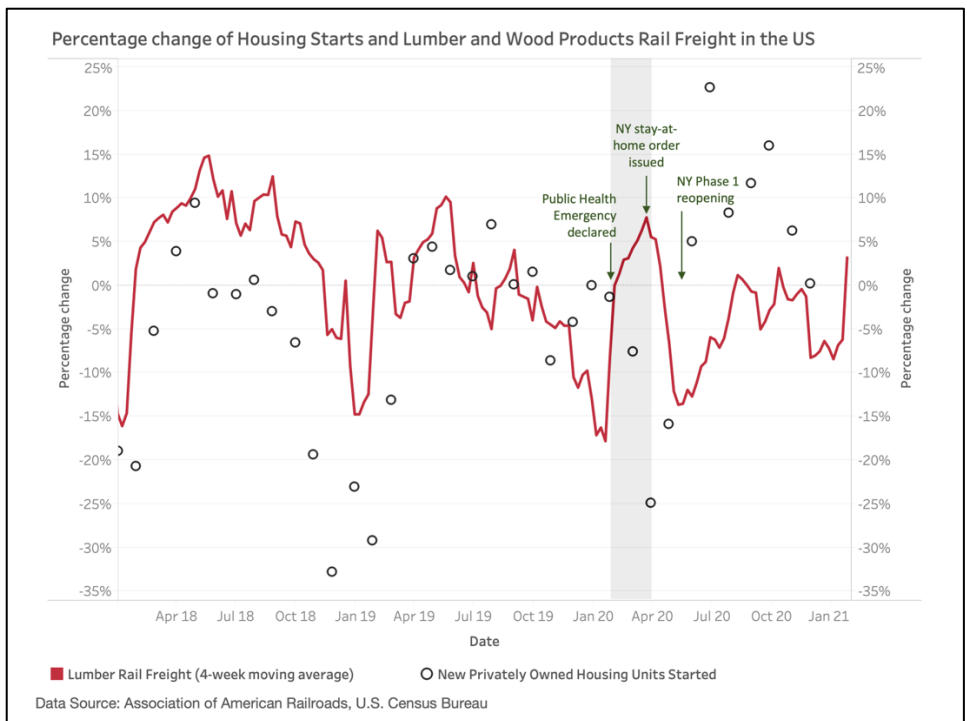

**Figure 26 - Percentage change of Housing Starts and Lumber and Wood Products Rail Freight in the US (2018-2021)**

**CONCLUSIONS**

This paper compares the changes in volumes of rail and intermodal freight before, during, and after the disruptions due to the Great Recession and the COVID-19 pandemic. These two most substantial rail disruptions in recent decades differed greatly in nature and characteristics. The Great Recession was a demand-side disruption led by an overall economic downturn, resulting in a more gradual yet long-lasting rail freight traffic decline and recovery. In contrast, the pandemic was a supply-side shock due to lockdown and other restrictions, manifested in the form of an abrupt drop followed by rapid recovery of rail freight.

The recovery of rail freight in 2020 was remarkable but uneven across markets. Intermodal was a strong driver of the quick rebound and continued growth under the new normal brought by the surge in demand for products over services. Products and commodities, such as motor vehicles and lumber, returned to pre-COVID levels, since the respective demands were either less affected or recovered promptly after the major lockdown was lifted and manufacturers adapted to the quick shift in consumer demand. A particular example was intermodal trailer freight associated with the rapid growth of e-commerce. This was largely different from the disruption to the rail freight industry during the Great Recession in 2007-2009, which was mainly driven by wide-ranging and lasting economic decline on both supply and demand sides.

The post-COVID return of coal and petroleum freight remained sluggish due to weak demand and low energy prices. Even though production was expected to rise after the pandemic, fierce competition with natural gas and transition to renewable energy appeared to limit their opportunity to attain previous peaks. They are also examples of rail freight market trends driven by secular development in other economic sectors.

Considering the shock brought by coronavirus in the second quarter of 2020, the rapid pace of recovery was a pleasant surprise for the economy, retailers that were able to adapt to new purchasing channels, and most consumers. A large part of sustained recovery and growth of intermodal rail freight is likely to be influenced by whether the long-term growth in e-commerce continues when the pandemic is over, and the extent to which intermodal freight will benefit from omnichannel retailing. (*18*)

By characterizing the fluctuations in freight volumes during the two disruptions, this study provides insights into patterns of freight market volatility driven by widely different factors, and illustrates the ways in which the logistics system, and particularly the U.S. railroads, responded. This analysis offers a perspective on the resilience of the freight transportation system and suggests the importance of growing that resilience to prepare for future disruptive events.

The analysis results presented here provide insights for various stakeholders in the supply chain to enhance their readiness for the next disruption. Policymakers can note the interaction between rail freight and general economy, improve their understanding of the factors that contribute to rapid response to unexpected economic disruptions, and look for new tools and incentives to promote a more resilience logistics system. Carriers (railroads, truckers) and shippers can better understand the roles and trends faced by other players in the supply chain, which might facilitate collaboration across the system. Researchers can extend the methods used here to study the resilience of the freight industry in the face of disruptive events and how it might be improved. In particular, there are opportunities to go beyond the macro-economic analyses presented here to capture a more detailed picture of cause and response to understand and evolve more resilient logistic systems.

## ACKNOWLEDGMENTS

The work presented here is based on work conducted at the Northwestern University Transportation Center (NUTC) (*19*). The authors are grateful to the provision of data and timely guidance from the Association of American Railroads and the contribution of Breton Johnson, NUTC Senior Associate Director, for facilitating industry contacts. The authors remain solely responsible for the content of this paper.

**AUTHOR CONTRIBUTIONS**

Max Ng conducted the data collection and analyses. All authors contributed to study design, analysis and interpretation of results, and manuscript preparation. All authors reviewed the results and approved the submission of the manuscript.

**DISCLAIMER**

The content of this paper is the sole responsibility of the authors and does not necessarily reflect the positions or policies of the Association of American Railroads or any of the organizations mentioned in this work.

## LIST OF FIGURES

**LIST OF TABLES**